\documentclass{aa}  

\usepackage{graphicx}
\usepackage{txfonts}

\usepackage{natbib}
\usepackage{amssymb}
\usepackage{color}
\usepackage{enumerate}
\usepackage{inputenc}

\newcommand {\aplt} {\ {\raise-.5ex\hbox{$\buildrel<\over\sim$}}\ }
\newcommand {\aco}[0] {{$\alpha_{\rm CO}$ }}

\begin{document}

\title{Modelling CO emission from hydrodynamic simulations of nearby spirals, starbursting mergers, and high-redshift galaxies}
\titlerunning{CO emission from galaxy simulations}

   \author{
   F. Bournaud
   \inst{1} \and
   E. Daddi
   \inst{1} \and
   A. Wei{\ss}
   \inst{2} \and
   F. Renaud
   \inst{1,3} \and
   C. Mastropietro
   \inst{1} \and
   R. Teyssier
   \inst{4} 
          }
\authorrunning{Bournaud et al.}

   \institute{Laboratoire AIM Paris-Saclay, CEA/IRFU/SAp, CNRS/INSU, Universit\'e Paris Diderot, 91191 Gif-sur-Yvette Cedex, France.
         \and
             Max-Planck-Institut f\"ur Radioastronomie (MPIfR), Auf dem H\"ugel 16, 53121 Bonn, Germany.
	\and
		Department of Physics, University of Surrey, Guildford GU2 7XH, UK.
         \and
             Institute for Computational Science, University of Zurich, CH-8057 Z\"urich, Switzerland.
             }

   \date{Received September 15, 1996; accepted March 16, 1997}

  \abstract{  We model the intensity of emission lines from the CO molecule, based on hydrodynamic simulations of spirals, mergers, and high-redshift galaxies with very high resolutions (3\,pc and $10^3$\,M$_\odot$) and detailed models for the phase-space structure of the interstellar gas including shock heating, stellar feedback processes and galactic winds. The simulations are analyzed with a Large Velocity Gradient (LVG) model to compute the local emission in various molecular lines in each resolution element, radiation transfer and opacity effects, and the intensity emerging from galaxies, to generate synthetic spectra for various transitions of the CO molecule. This model reproduces the known properties of CO spectra and CO-to-H$_2$ conversion factors in nearby spirals and starbursting major mergers. The high excitation of CO lines in mergers is dominated by an excess of high-density gas, and the high turbulent velocities and compression that create this dense gas excess result in broad linewidths and low CO intensity-to-H$_2$ mass ratios. When applied to high-redshift gas-rich disks galaxies, the same model predicts that their CO-to-H$_2$ conversion factor is almost as high as in nearby spirals, and much higher than in starbursting mergers. High-redshift disk galaxies contain giant star-forming clumps that host a high-excitation component associated to gas warmed by the spatially-concentrated stellar feedback sources, although CO(1-0) to CO(3-2) emission is overall dominated by low-excitation gas around the densest clumps. These results overall highlight a strong dependence of CO excitation and the CO-to-H$_2$ conversion factor on galaxy type, even at similar star formation rates or densities. The underlying processes are driven by the interstellar medium structure and turbulence and its response to stellar feedback, which depend on global galaxy structure and in turn impact the CO emission properties.}
   \keywords{Galaxies: ISM, star formation --- interstellar medium: structure, clouds, molecules}
   
   \maketitle
   
%

\section{Introduction}
Understanding the molecular gas content of galaxies at various stages of their formation is an increasingly important issue in the field of galaxy evolution, especially with high-resolution campaigns using radio interferometers \citep[e.g.,][]{daddi-co, tacconi-co, tacconi-co2, combes-co, sargent, walter-co, paws, paws2}. In parallel, the last generations of cosmological simulations start resolving sub-structures in the interstellar gas reservoirs of galaxies with gas densities about up to that of star-forming molecular clouds \citep{ceverino12, hopkins}. Interpreting observations of the molecular tracers of the ISM, atop of which the CO molecule, and comparing with theoretical predictions on the bulk of the molecular mass (i.e., the H$_2$ molecule), remains highly uncertain. The variations in the excitation of the CO emission, and the CO-to-H$_2$ conversion factor, are indeed very large, and their interpretation largely relies on arbitrary calibrations. The conversion factors often used to estimate the H$_2$ mass depend on various factors, such as the metallicity \citep{Israel, genzel-z}. They also appear to vary with galaxy type in the nearby Universe \citep{solomon1,solomon2}, which led to the observation that various types of galaxies may also form their stars with very different efficiencies (\citealt{daddi-ks,genzel-ks}) -- a result questioned for being dependent on arbitrary assumptions on \aco (see e.g. \citealt{feldman-ks}, \citealt{ppdpl2} and \citealt{sargent} for detailed discussions).

An increasing amount of theoretical work is performed with analytic and numerical models, in particular to implement the H$_2$ molecule formation in high-resolution simulations and/or to model the emission of the CO molecule \citep[e.g.,][]{narayanan1,narayanan2,narayanan3,krumholz, feldman1,feldman2, ppdpl,nara-krumholz}. In this paper, we introduce a novel approach combining Large Velocity Gradients (LVG) analysis with high-resolution Adaptive Mesh Refinement (AMR) simulations of various types of galaxies: these simulations ideally describe the ISM phase-space structure on small scales (down to a few pc) and up to very high densities ($\geq 10^6$\,cm$^{-3}$), and LVG post-processing determines the emergent intensity of CO emission lines from this ISM structure. The underlying assumption of this technique is that opacity effects occur mostly on relatively small scales (a few parsecs), because the low volume filling factor of the molecular ISM makes unlikely for two clouds to be found on a given line-of-sight with co-incident Doppler shifts to allow for large-scale re-absorption. On small-scales, depending on the (turbulent) kinematic structure of the clouds, the interstellar medium becomes transparent to its own emission owing to Doppler shifting of the molecular lines beyond a sufficient distance -- generally a few pc or less, as determined from the kinematic structure of the AMR simulations coupled to the LVG model grids. These grids also determine the CO emissivity in various lines as a function of temperature, density, metallicity and redshift (CMB temperature).

We use this technique to analyze and compare simulations of various types of galaxies: nearby spiral galaxies with low gas fractions and secularly-evolving disks, high-redshift disk galaxies with high gas fractions and disk instabilities, and interaction/merging galaxies in a starburst (SB) phase, i.e. galaxies that have recently undergone a major increase of their star formation rate under the effect of a major interaction/merger (not all galaxy merger undergo such phases and these may last shorter than the merging process itself, \citealt{dimatteo07, ellison}). We probe the effect of galaxy type and structure on the CO excitation and \aco conversion factor, and to identify these effects we fix some other physical parameters, in particular the metallicity is fixed at solar value (relatively massive galaxies at redshift $z$=1-2 are already significantly enriched anyways -- \citealt{erb,queyrel,ewuyts}). We find in particular that high-redshift disks, although they are very gas rich and unstable, have global CO properties that are relatively similar to Milky Way-like spirals, however with locally substantial high-excitation components in their giant star-forming clumps. SB mergers, even though they form stars about at the same rate as high-redshift disks, have very different CO properties arising from a different small-scale ISM structure. 

The simulations used in this work and the analysis technique are presented in Section~2. The results obtained for various types of galaxies are analyzed in Section~3, and they are compared to other theoretical and numerical models in Section~4. 
Throughout the manuscript, the CO luminosity-to-H${_2}$ mass conversion factor \aco is given in $\mathrm{M_\odot\,(K\, km \, s^{-1} \, pc ^2)^{-1}}$.


\section{Simulations and Analysis}

\begin{table*}
\caption{ \label{table-models} Physical properties of the three simulations representative of various galaxy types in this work. }
\centering
\begin{tabular}{lcccc}
\hline\hline
Model & $M_*$ & $f_{\rm gas}$ & Redshift & $<$sSFR$>$ \\
\hline
spiral  & $6.5 \times 10^{10}$\,$M_\odot$                               & 9\% & $z = 0$ & 0.04\,Gyr$^{-1}$ \\
SB merger & $6.5 \times 10^{10}$\,$M_\odot$ ($\times$2) & 9\% & $z = 0$ & 0.63\,Gyr$^{-1}$ \\
high-z disk  & $3.5 \times 10^{10}$\,$M_\odot$                     & 60\% & $z = 2$ &  0.97\,Gyr$^{-1}$ \\
\hline
\end{tabular}
\tablefoot{The specific star formation rate (sSFR) indicated here is averaged over the snapshots analyzed for each galaxy type: random for the spiral and high-z disk model, picked only in starbursting phases for the SB merger model. The stellar mass and gas fraction are the initial parameters of the models. A representative redshift is indicated for each model, it is also used as an input parameter for the CMB temperature in the LVG grids. }
\end{table*}

\begin{figure}
\centering
\includegraphics[width=0.45\textwidth]{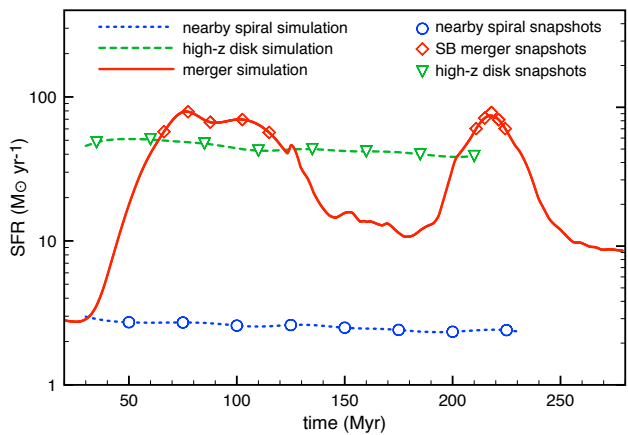}
\caption{\label{fig_SFR} \small 
Star formation rate as a function of time for the three galaxy models analyzed in this work. The snapshots/instants analyzed for each galaxy type are highlighted with various symbols: they are chosen randomly (about evenly spaced in time) for disks, and only in starburst phases with elevated SFR for the major merger simulation. The corresponding sSFRs (averaged for each type) are given in Table~\ref{table-models}.}
\end{figure}

\subsection{Simulation Sample}

The three simulations used here are originally provided by different studies, but were performed with the same Adaptive Mesh Refinement (AMR) code RAMSES \citep{teyssier02}, and use similar spatial resolutions of 3.0 to 3.3\,pc and similar AMR refinement schemes, and including the same physical ingredients : gas cooling at solar metallicity, heating with a ultraviolet background\footnote{We here used a standard UV background representative for the Galaxy and ow-redshift conditions. Redshift evolution of the UV background is neglected here, but likely to affect mostly the low-density phases of the ISM, rather than the molecular clouds that are mostly shielded from the UV background.}, and comprehensive modeling of stellar feedback processes from \citet{Renaud13} including photo-ionization, radiation pressure, and type~II supernovae. A complete description of the simulation methods can be found in \citet{Bournaud14} or \citet{Renaud14}.

We use three high-resolution simulations, representative of high-redshift disk galaxies, low-redshift spiral galaxies, and low-redshift starbursting mergers, all starting with similar baryonic masses of 6--$7\times10^{10}\,M_\odot$: 
\begin{itemize}
\item a {\em spiral} representative of low-redshift disk galaxies, with a gas fraction of 9\% of the baryonic mass, and a morphology dominated by spiral arms and small gas clouds forming mostly along the spiral arms. The star formation rate is 2--$3\,M\odot\,yr^{-1}$. This is the control disk run from Renaud et al. (2014a,b). 

\item a {\em high-z disk} model with high gas fraction (initially 60\% of the baryonic mass, decreasing to 50--40\% along the simulation through gas consumption), subject to violent disk instability and spontaneously taking an irregular morphology dominated by young star-forming clumps, with high turbulent velocity dispersions and significant outflows, all consistent with observations: this is model G'2 from \citet{Bournaud14} and \citet{Perret14}, re-computed here at 1.7\,pc resolution for consistency with the low-redshift simulations. The star formation rate is consistent with the ``Main Sequence'' of star-forming galaxies at $z \, \approx \, 2$.

\item  a {\em starbursting (SB) merger}, produced by a major merger between two spiral galaxies similar to the {\em spiral} model used here. The chosen merger model, from Renaud et al. (2014a,b) has an orbit and a tidal field quite representative for mergers in $\Lambda$-CDM cosmology. The interaction triggers starburst events peaking at 80--$100\,M\odot\,yr^{-1}$ at the first pericentric encounter and during the final coalescence : the analyzed snapshots correspond to these phases when the merging system is actually starbursting about a strong LIRG or moderate ULIRG-like level, with an SFR\footnote{averaged over 10\,Myr} in the 45--100$M\odot\,yr^{-1}$ range. Phases with more quiescent star formation are not considered as representative of an SB merger and ignored in the present analysis. 
\end{itemize}

Table~1 summarizes the key properties of each simulation. The star formation rate evolution in each one is displayed in Figure~\ref{fig_SFR}. {\bf In this paper we explore the properties of CO emission as a function of galaxy type, rather than as a function of metallicity (see also discussion in Section~4). Hence we assume solar metallicity for all cases, since massive star-forming disks at $z \approx 2$ are already relatively metal-rich \citep{erb,queyrel-z}. The spatial structure of the ISM and the star formation efficiency can be affected by metallicity, but only much lower metallicities (a tenth solar and below) would have a strong impact \citep{kraljic14}. The CO abundance in molecular gas also depends on metallicity, but for the average metallicity variations between $z$=0 and 2 for massive disk galaxies (less then a factor two) the metallicity effect on \aco remains modest \citep{leroy11,genzel-z} and eventually not larger than the effect of galaxy type found in our present results. }

Note that the SB merger and spiral simulation exist at 1.5\,pc resolution (Renaud et al. 2014) but such extremely high spatial resolution is hard to reach for the high-z disk model without lowering the mass resolution, because the higher gas mass in this case implies a larger number of resolution elements. For consistency we here analyse all models at 3\,pc resolution, knowing that the star formation rate history and density distribution (PDF) have been shown to converge with respect to resolution at this 3\,pc scale in Renaud et al. (2014).
 
 \medskip
It is known from previous analysis of these simulations (Bournaud et al. 2014, Renaud et al. 2014, Kraljic et al. 2014) that these models have a global star formation efficiency (or gas depletion timescale) in agreement with the observed Schmidt-Kennicutt relation(s) (Kraljic et al. 2014, Renaud et al. 2014) and that the high-z disk models proceed giant clumps of gas and star formation from which intense outflows are launched (the giant clumps in these high-z galaxies typically contain 20-35\% of the gas mass, up to $\sim$50\% of the SFR, and a few percent of the total stellar mass, Bournaud et al. 2014). It is also known that the ISM is more turbulent in both high-z disks and SB mergers than in spirals, with typical turbulent velocity dispersions in the 20-50\,km\,s$^{-1}$ range for high-z disks and mergers, compared to 5-10\,km\,s$^{-1}$ for spirals. In addition, turbulent motions have been shown to be preferentially compressive in SB mergers compared to spirals and high-z disks: this means that around density peaks, radial turbulent motions compressing (or disrupting) the density peak are favored w.r.t. solenoidal motions curling around the density peaking, in the case of galaxy mergers (Renaud et al. 2014).

\medskip

Each simulation was analyzed at different snapshots, regularly spaced in time for disks, and in the starbursting phases for the SB merger model, as also indicated in Figure~\ref{fig_SFR}. For disks, each snapshot was analyzed under three lines-of-sight: one face-on orientation, and two projections inclined by 60 degrees. The SB merger snapshots were analyzed under three perpendicular lines of sight: in the following, the average result was used to compute the molecular line emission for each galaxy model, unless stated otherwise.

\subsection{LVG analysis}

The interstellar gas distribution in AMR simulations is described with cells of variable size. Each cell has a size $\epsilon$, and contains gas at a density $\rho$, temperature $T$, velocity $\vec{v}$, metallicity $Z$, etc. In the present simulations, the cell size $\epsilon$ is in the 3--12\,pc range for gas denser than 10\,cm$^{-3}$, coarser cells correspond only to low-density (atomic or ionized) gas.

The purpose of the LVG analysis is to estimate the flux emerging from each individual cell for any given CO($J$-$J-1$) transition, and assuming that the line-of-sight of the pseudo-observation is given by the unitary vector $\vec{u}$. In the LVG framework, the raw emission is determined by the physical parameter of the gas in the cell, but the radiation can be re-absorbed by molecules in the same cell, i.e. in the immediate (parsec-scale) vicinity of the emitting regions. This local re-absorption depends on how rapidly the line frequency shifts along the line of sight : if strong velocity gradients are present, the line central frequency is shifted by an amount larger than the intrinsic linewidth over a short distance and the gas becomes transparent to its own emission, while weak velocity gradients imply that gas can re-absorb its own emission over larger distances. To account for this we compute the $dV/dr$ parameter, where $r$ is the position along the line of sight axis $\vec{u}$, given by $dV/dr  =\sum \limits_{i=1,2,3} {u_i \partial V_i / \partial r_i}$. The gradient $\partial V_i / \partial r_i$ is determined on each axis $i$ by comparing the velocity field $\vec{v}$ over neighboring cells.

The simulations employ a temperature threshold at high densities to prevent artificial fragmentation effects \citep[][Section~2.1]{Teyssier10}: the hydrodynamic equations are solved assuming a pressure that cannot be smaller than a value calculated to keep the Jeans length larger than four cells at the maximal refinement level. {\bf The corresponding ``numerical'' temperature thus cannot be smaller than 50--100\,K in our simulations.} However, the extra numerical pressure can be considered as a sub-grid resolved for non-thermal processes (such as small-scale turbulent motions and stellar feedback winds rather than thermal energy, e.g., Teyssier et al. 2010). The pressure floor is then subtracted during the analysis to obtain only the physical temperature, i.e., the amount of thermal energy heating the gas above the menial temperature floor. {\bf We hence resolve cold high-density components down to 10-20\,K, where the temperature here represents the true thermal energy after subtracting the extra numerical components (see above). Note that cooling through rotational molecular transitions in the molecular phase is not included in our simulations, yet a two-phase interstellar medium is produced with a density-temperature distribution (Perret et al. 2014, Fig. 1)consistent with more detailed simulations of the ISM \citep[e.g., ][]{kim}}

The combination of the local gas physical parameters and the $dV/dr$ measurement of the local, small-scale opacity of the ISM to its own emission directly indicates the net flux emerging in each CO(J--J-1) transition through the LVG tables from \citet{weiss05}, assuming collision rates from \citet{flower} and a standard CO abundance \citep{frerking,lacy}. The underlying assumption in this modeling is that the CO emission is not re-absorbed on large scales that are well resolved in our simulations ($\leq$10-100\,pc), because the turbulent structure of the molecular ISM makes it unlikely to find two molecular clouds along the same line-of-sight and with a velocity difference small enough to allow an overlap of the CO emission/absorption lines with their intrinsic linewidth (a few km\,s$^{-1}$, lower than the velocity shifts induced by galactic rotation on large scales). The same technique can be used for other molecular transitions such has HCN lines, not studied in the present paper. 
\medskip

Some of the calculations presented hereafter require to estimate the total molecular mass (i.e., the $H_2$ mass). Our numerical simulations are performed at very high resolution, which has the advantage to explicitly resolve the dense gas clouds and their substructures, up to densities larger than $10^6$\,cm$^{-3}$, without requiring to assume a sub-grid model for the cloud mass function during the post-processing. This however comes at the expanse of the inclusion of a molecule formation model in the simulations themselves, as could be done with a reasonable computer cost in simulations employing more modest maximal resolutions \citep{narayanan1,feldman1}. {\bf When required, we hereafter assume that the ISM becomes molecular above a simple density threshold. We use two values for the threshold, of 10 and 50\,cm$^{-3}$, and keep the average result: these two values bracket the typical densities at which the ISM starts to be molecule-dominated in state-of-the-art models of entire galaxies with accurate molecule formation scheme \citep[e.g.][]{feldman2}. The fraction of the ISM mass encompassed in the 10-50\,cm$^{-3}$ range is small enough for the related uncertainty on the results to be limited, and smaller than case-to-case fluctuations for each type of galaxies (uncertainty below 10\% for the \aco). }

\section{Results}

\subsection{CO spectral line energy distributions}

\begin{figure}
\centering
\includegraphics[width=0.47\textwidth]{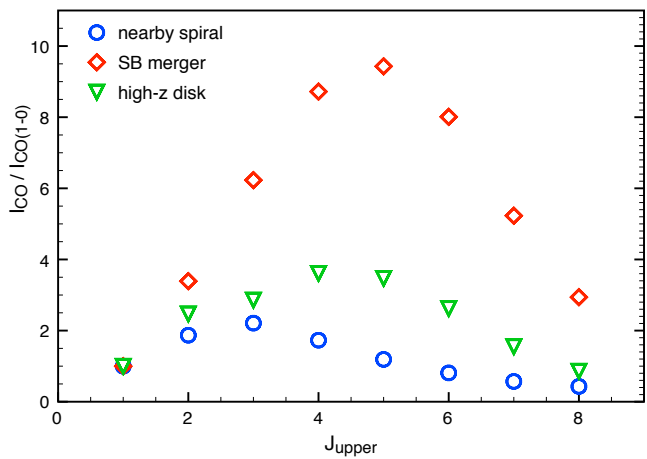}
\caption{\label{fig-sled} \small 
Spectral Line Energy Distributions (SLEDs) of the CO molecular for the three galaxy types, averaged over the different snapshots and projections analyzed for each galaxy type. The SLEDs are normalized to the CO(1-0) line intensity. See text for details on the fluctuations between snapshots and projections at fixed galaxy type.}
\end{figure}

The CO spectral line energy distributions (SLED) are shown in Figure~\ref{fig-sled} for each type of galaxy, averaged over the different snapshots and projections analyzed, and normalized to the CO(1-0) line intensity. While there are strong differences with galaxy type, the differences between different snapshots and/or projections for a given galaxy type are relatively small. The rms individual variations of the $I_{CO(2-1)}/I_{CO(1-0)}$ over individual snapshots range from 12\% for the spiral galaxy to 16\% for the high-z disk and 32\% for the SB merger. The rms variations of the $I_{CO(5-4)}/I_{CO(1-0)}$ range from 21\% for the spiral galaxy to 27\% for the high-z disk and 56\% for the SB merger. Hence the differences in the average CO SLED between the various types of galaxies, as seen on Figure~\ref{fig-sled}, are much larger than the intrinsic variations at fixed galaxy type. 

Note that the intrinsic variations of the SLED are larger for the SB merger than for the spiral and high-z disk, which can be attributed to a varying level in the starburst activity: all the selected snapshots for the SB merger correspond to phases where the SFR is elevated, but the exact level of SF activity varies, and the physical properties of the starbursting ISM also vary in this model. The starburst starts in a highly fragmented, spatially-extended medium, and evolves toward a more concentrated nuclear starburst, hence qualitatively recovering the observed variety of starbursting mergers (see Renaud et al. 2014 for details). Based on this, and given that the SB merger analyzed here is about the ILRG/ULIRG transition, its average SLED should be considered as representative for such systems, relatively common among major mergers. More extreme starbursts that are less common at least at low redshift could likely show even larger excitation in their CO SLED properties.

The CO SLED of the spiral galaxy model is consistent with that of the Milky Way disk, and that of the SB merger model is consistent with observations of starbursting mergers about the LIRG/ULIRG activity levels (see Introduction). Our AMR simulations and LVG analysis thus yield realistic results for these two class of galaxies, for which the typical SLED properties are relatively well constrained observationally -- as opposed to high-redshift disks. As for high-redshift disk, we find that the predicted CO SLED is overall closer to that of a spiral galaxy than of a starbursting merger, but with a significant excess of high-excitation components compared to spiral galaxies. The  $I_{CO(2-1)}/I_{CO(1-0)}$ and $I_{CO(3-2)}/I_{CO(1-0)}$ ratios for the high-z disk are moderately larger than for the spiral galaxy model, but these ratios become much larger than in the spiral galaxy model when CO transitions with $J_{upper} \geq 4$ are considered. Actually, the high-z disk model present a surprisingly high $I_{CO(4-3)}/I_{CO(3-2)}$ ratio, larger than its $I_{CO(3-2)}/I_{CO(2-1)}$ ratio, a property that is not met in spiral disks and SB mergers. This property is found in almost all snapshots (7 out of 8) of the high-z disk model. The physical origin of the high-excitation components in high-z disks compared to SB mergers will be analyzed later in Section~4.


\subsection{Dissecting high-redshift clumpy disks}

\begin{figure}
\centering
\includegraphics[width=0.5\textwidth]{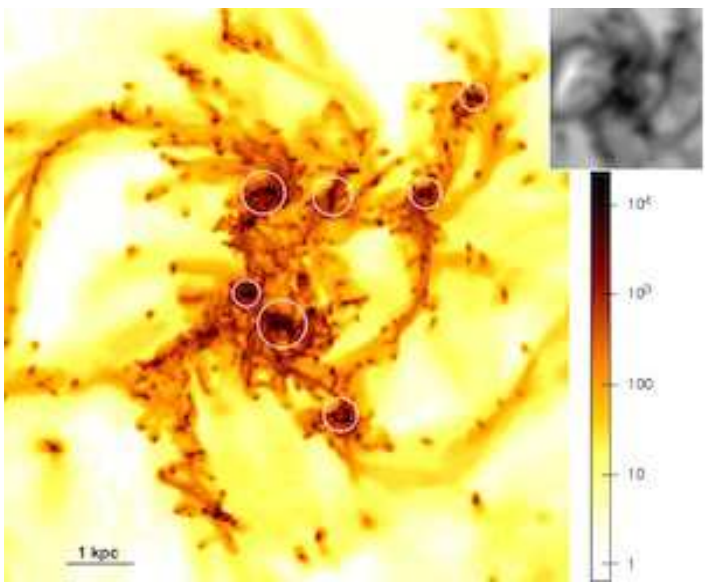}
\caption{\label{fig-snapshot} Face-on gas density map of a snapshot of our high-z disk model, used to analyze the contribution of ``giant clumps'' to the global CO SLED. The gray-scale inset displays a mock HST-like optical observation (B-band rest-frame emission with a gaussian PSF of FWHM 500~pc), showing an irregular structure with a few giant clumps and some spiral arms, typical for real $z \approx 2$ star-forming galaxies (Elmegreen et al. 2007). The seven giant clumps (with stellar masses above $10^8\,M_\odot$) are circled on the gas map, with the clump radius computed as in Bournaud et al. (2014). These seven clumps contain about half of the total SFR and one third of the total gas mass. The SLEDs corresponding to the emission for the clump regions and from the rest of the galaxy are analyzed in Figure~\ref{fig-sled-deco}. Gas number densities are in cm$^{-3}$, the map is smoothed at a 10\,pc scale.}
\end{figure}

\begin{figure}
\centering
\includegraphics[width=0.47\textwidth]{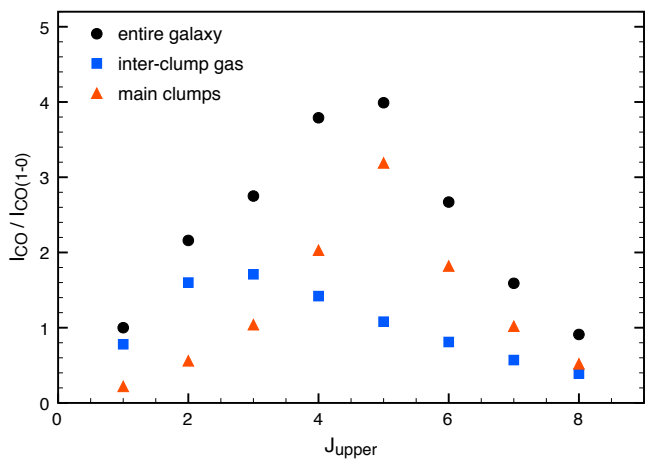}
\caption{\label{fig-sled-deco} CO SLEDs for the total galaxy, the seven giant clumps, and the rest of the gas, on the high-z disk snapshot shown in Figure~\ref{fig-snapshot}.}
\end{figure}

To understand the origin of the high-excitation components in the CO SLED of high-redshift disks compared to nearby spirals, we perform a spatially-resolved analysis of a representative snapshot shown on Figure~\ref{fig-snapshot}. This is the third snapshot in the time sequence of our high-z disk simulation, so it has a gas fraction and clumpiness close to the average of the other snapshots and is representative of a typical $z \approx 2$ main sequence disk galaxy. We checked separately that the other snapshots show the same behavior in their spatially-resolved properties, and here we show the results averaged over three perpendicular projections for the selected snapshots. 

The chosen high-redshift disk snapshot has a clumpy and irregular morphology, typical for high-redshift galaxies, and a gas fraction of 53\% at the instant analyzed. We identify the 7 most massive clumps and measure their optical radius as in Bournaud et al. (2014). These giant clumps have baryonic masses in the 1.7--7.3$\times$$10^8$\,M$_\odot$ range, and contribute to 57\% of the total SFR of the galaxy. The inter-clump medium has a lower average density but does contain high-density clouds on small scales, with individual masses and sizes closer to that of low-redshift molecular clouds. This provides a significant component of relatively diffuse star formation (although clustered on small scales $<$100\,pc).

The CO SLEDs were measured separately for the main giant clumps (summed over the clumps), and for the rest of the galaxy, i.e. the inter-clump medium. The results are shown and compared to the entire galaxy on Figure~\ref{fig-sled-deco}. The clumps gas has a high-excitation SLED with an intensity ratio peaking at the 5-4 transition, qualitatively similar to the starbursting merger type. At the opposite the SLED of the inter-clump medium peaks in the CO(3-2) line, and the relative line ratios are close to those measured for nearby spirals, with differences smaller than 20\% for CO(2-1), CO(3-2) and CO(4-3) intensities relative to CO(1-0). A moderate excess is found for higher-J transitions (about 35\% in intensity) compared to nearby spirals.

The global CO SLED of high-z star-forming disk galaxies thus appears to result from the combination of a low-excitation, about Milky Way-like component in the diffuse disk, containing about half of the total gas mass and one third of the star formation rate, and a higher-excitation component in a few giant clumps containing two third of the total star formation rate. The origin of the high CO excitation therein will be analyzed in the next section. The combination of these two components results in the increased CO(4-3)/CO(3-2) line intensity ratio compared to the CO(3-2)/CO(2-1) one, a feature found in the high-z snapshot analyzed here (Fig.~\ref{fig-snapshot}, \ref{fig-sled-deco}) and on average for high-z disks snapshots (Fig.~\ref{fig-sled}).

\subsection{CO luminosity-to-molecular gas mass conversion factor}

\begin{table}
\caption{ \label{table_aco} Average value of the \aco conversion factor for each galaxy type, averaged over all snapshots and projections. The indicated range corresponds to the extreme assumptions for the molecular mass definition. $\alpha_{\rm CO}$ is in $M_\odot\,(K\, km \, s^{-1} \, pc^2)^{-1}$ throughout the manuscript.}
\centering
\begin{tabular}{lcc}
\hline\hline
Model & & $\alpha_{\rm CO}$ range    \\
\hline
$\,\,\,\,$ spiral           & $\,\,\,\,\,\,\,\,$ &  4.3 -- 4.7  $\,\,\,\,\,\,\,\,$ \\
$\,\,\,\,$ high-z disk  &                & 3.9-- 4.5 $\,\,\,\,\,\,\,\,$  \\
$\,\,\,\,$ SB merger    &                & 1.7 -- 2.2 $\,\,\,\,\,\,\,\,$\\
\hline
\end{tabular}
\end{table}

\begin{figure}
\centering
\includegraphics[width=0.35\textwidth]{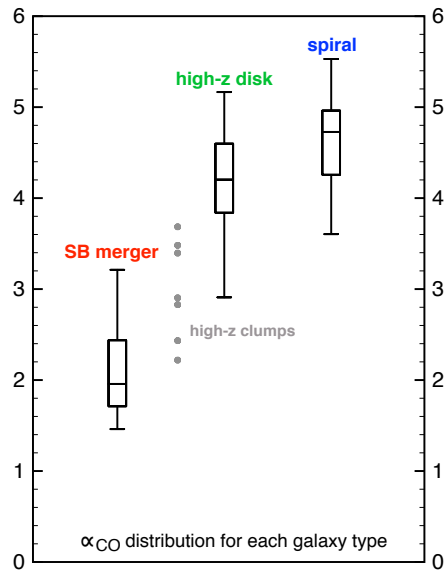}
\caption{\label{fig_aco} \small 
Statistical distribution of the CO luminosity-to-H$_2$ mass conversion factor (\aco) for each galaxy type. The symbols indicate the extreme values, quartiles, and median for each galaxy type. The statistical scatter is dominated by variations between snapshots and projections, the uncertainty on the molecular mass itself being a minor contribution to the scatter (see text). The gray dots show individual giant clumps in a representative high-redshift galaxy (see Section~3.2 and Fig.~3).}
\end{figure}

The CO(1-0) line intensity measured in the various snapshots and projections is compared to the mass of molecular gas present in the simulations, derived with various assumptions detailed in Section~2. This yields a statistical distribution of the \aco conversion factor for each galaxy class, displayed in Figure~\ref{fig_aco}.  We summarize in Table~\ref{table_aco} the average value of \aco for the highest and lowest assumptions on the molecular gas mass, for each galaxy type. Our extreme assumptions on the molecular gas mass definition yield an uncertainty of 6--10\% on the average \aco for the various galaxy types. This is smaller than the variations between snapshots and projections, which induce an r.m.s. variation in the recovered \aco factor of 11\% for spirals, 13\% of high-redshift disks, and 22\% for starbursting mergers. 

\begin{figure}
\centering
\includegraphics[width=0.47\textwidth]{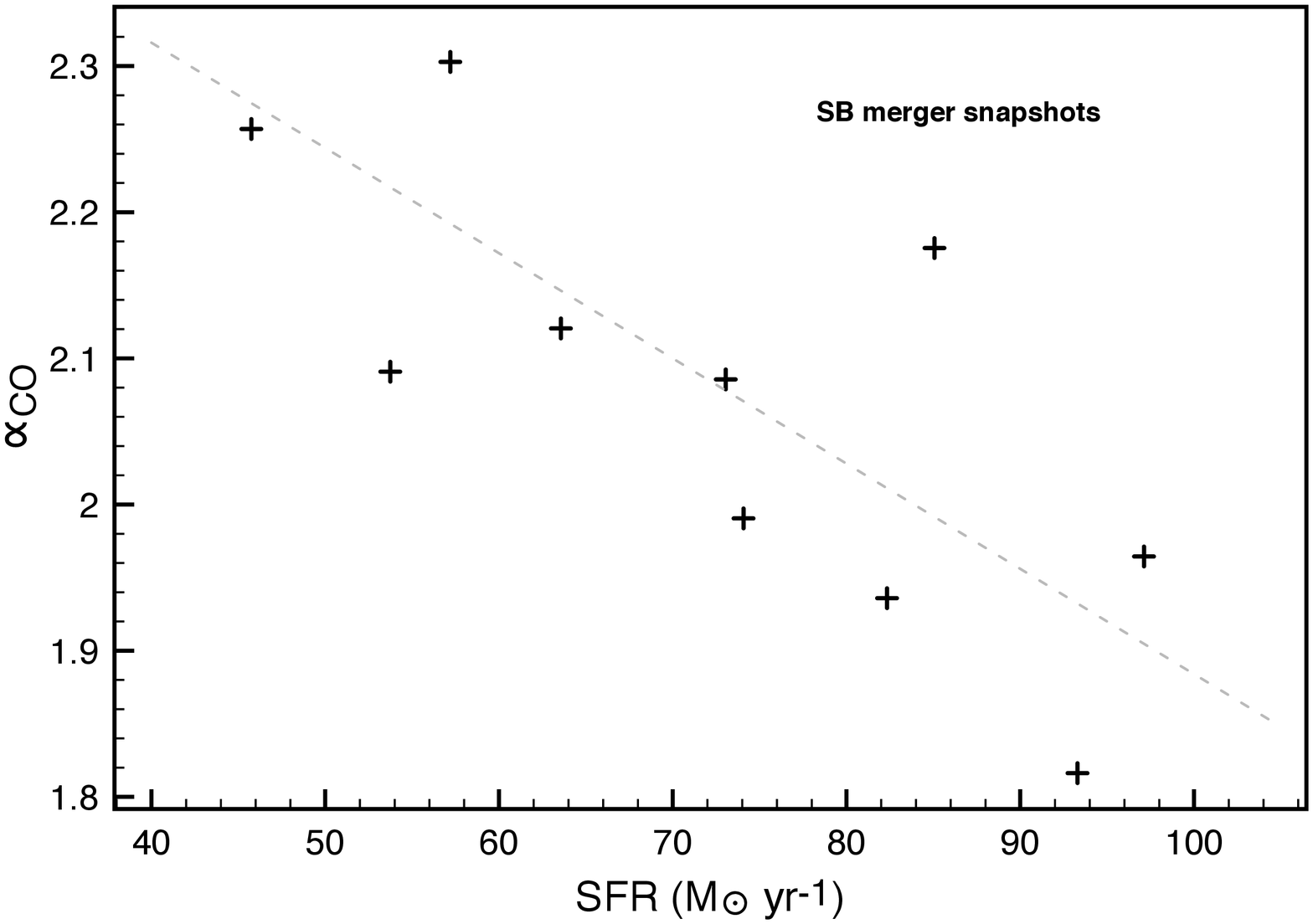}\\
\includegraphics[width=0.47\textwidth]{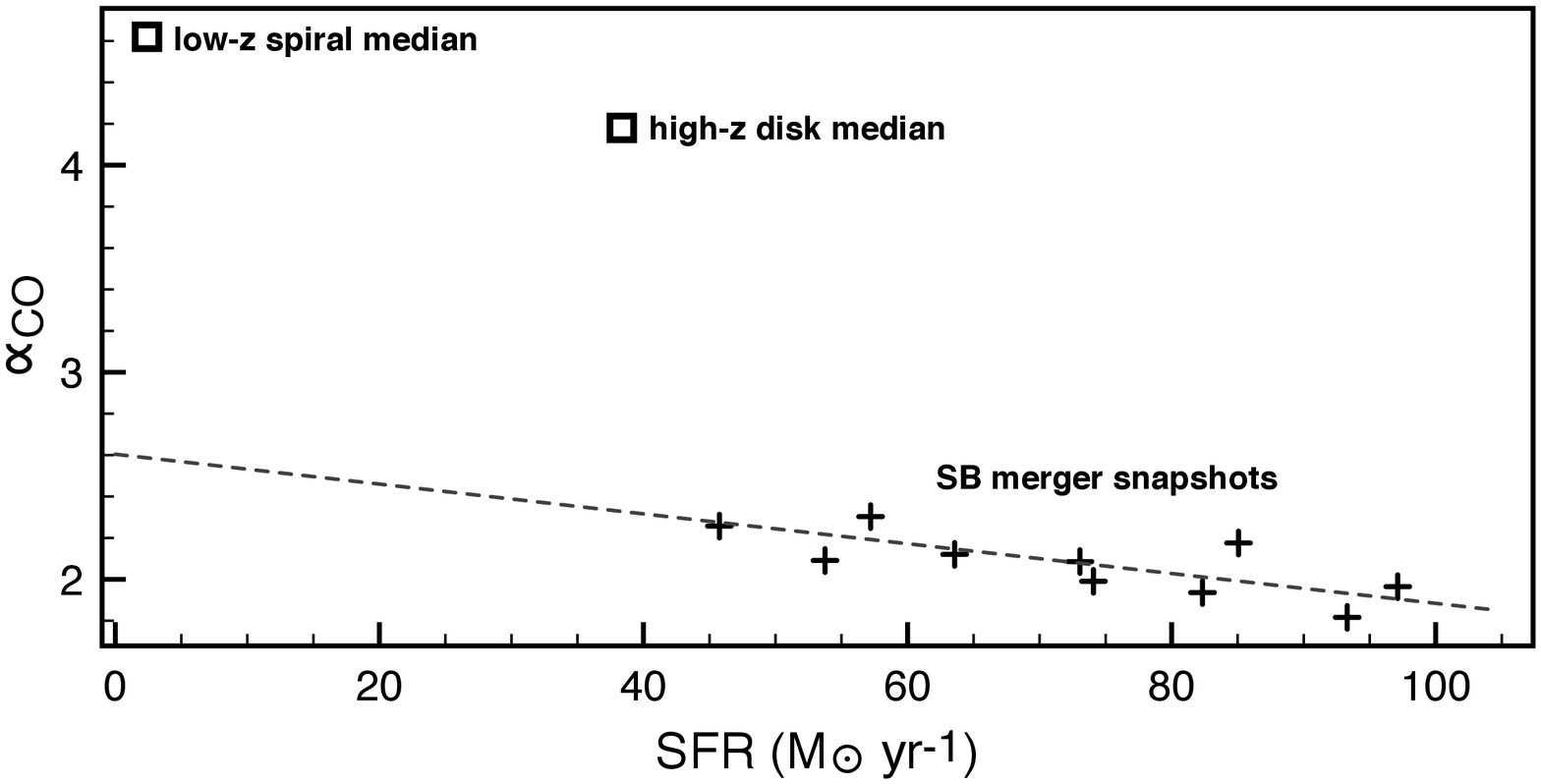}
\caption{\label{fig_aco_sfr} Comparison of the retrived \aco conversion factor to the SFR for each individual snapshot of the SB merger model (top -- averaged over three projections for each snapshot). A clear correlation is found between SFR and \aco within the SB merger galaxy type, and the intrinsic variations of the SFR in the SB merger model induce a larger scatter in \aco than for isolated disk models. This indicates that stronger starbursts could have even lower \aco factors, given that the merger model used here is moderately starbursting about the LIRG/ULIRG transition regime. In contrast, other galaxy types do not follow this \aco--SFR trend retrieved for the SB merger snapshots (bottom). In particular, high-z disks have an SFR almost as high as some SB merger snapshots, but with a much higher \aco (see also Fig.~\ref{fig_compar}).}
\end{figure}

The larger relative variations of \aco for SB mergers correspond to the natural variations of the SFR activity along the merger time sequence. We compare in Figure~\ref{fig_aco_sfr} the median estimate of \aco to the SFR for individual snapshots of the SB merger model. There is a trend for lower \aco values when the SFR is the highest, i.e. about 80-100\,$M\odot\,yr^{-1}$. While our SB merger is just reaching about the level of ULIRGs, this suggests that more extreme merger-triggered starbursts likely have even lower \aco values (although higher SFR enhancements are rare, see e.g. \citet{dimatteo07}, so higher star formation rates in mergers may require different type of progenitor galaxies).

However, when considering the various systems in our sample (low-$z$ spiral, high-$z$ disk, SB merger), there is no uniform trend between the \aco factor and the SFR, and the disks lie far off the trend retrieved between SB merger snapshots (Fig.~\ref{fig_aco_sfr} bottom). Actually, our high-z disk model has an SFR close to the most moderate SB merger snapshots selected for our analysis, but has a much higher \aco. We will show below that this relates to different small-scale phase-space structures of the ISM. Such a behavior of \aco, only weakly varying for Main Sequence galaxies independent of redshift and SFR, but rapidly dropping for starburst systems with an $\alpha_{\rm co}$--SFR correlation for the latter category, is consistent with the results of \citet[][Fig.~15b]{sargent14} (in detail, Sargent et al. find a somewhat stepper relation between \aco and SFR for starbursts, but with a broad scatter).

We also show in Figure~5 the \aco factor for the giant clumps identified in a representative snapshot of the high-z disk model (the one used in Figs.~3, 4 and Sect.~3.2). These giant clumps have individual \aco factors lower than their entire host galaxy, although not as low as the SB merger model. This appears consistent with the fact that these kpc-clumps behave like moderate starburst regions in the Schmidt-Kennicutt diagram, namely they have gas depletion timescales shorter than their entire host galaxies, but not as short as the most intensely starbursting mergers (\citealt{freundlich} for observations, Zanella et al. in preparation for simulations). The low \aco factors of the clumps explain why the \aco of entire high-z disks is somewhat lower than that of spirals: on average 23\% of the CO(1-0) line intensity from the whole galaxy is emitted by the highly-excited giant clumps, while the bulk of the CO(1-0) emission remains dominated by the low-excitation large-scale disk.

\section{Interpretation}

\subsection{Origin of the high-excitation components in starbursting mergers and high-redshift disks}

Here we analyze further the origin of the strong excitation of high-J CO lines, first in SB mergers compared to spirals, then in high-z disks (in particular their giant star-forming clumps). In particular, we aim at disentangling the role of high-density gas and warmed gas in the CO line properties of such systems.

\subsubsection{Starbursting mergers}

\begin{figure}
\centering
\includegraphics[width=0.5\textwidth]{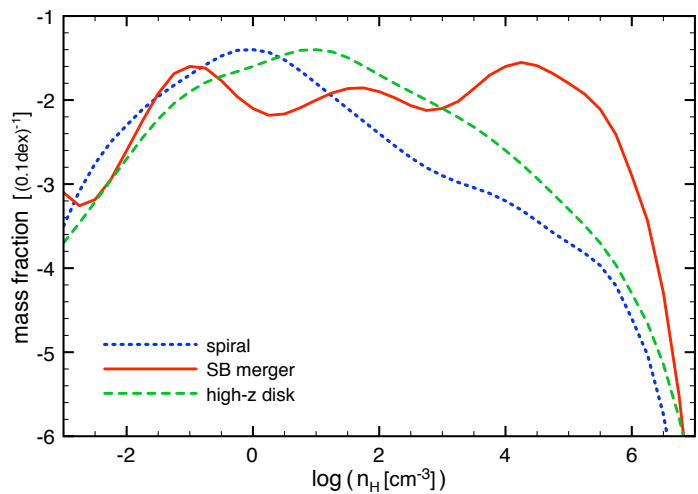}
\caption{\label{fig-pdf} Density point distribution function (PDF) of the gas for each galaxy type (averaged over the different snapshots). The local spiral and high-z disk models have density PDFs that are well described by a log-normal functional shape, with a small excess of high-density gas in the form of a power-law tail, which corresponds to self-gravitating gas \citep{Renaud13, Bournaud14}. SB mergers strongly deviate for a log-normal PDF and have a strong excess of high-density gas. The rapid drop of the PDFs above $10^6$~cm$^{-3}$ corresponds to the resolution limit of our simulations.}
\end{figure}

\begin{figure}
\centering
\includegraphics[width=0.5\textwidth]{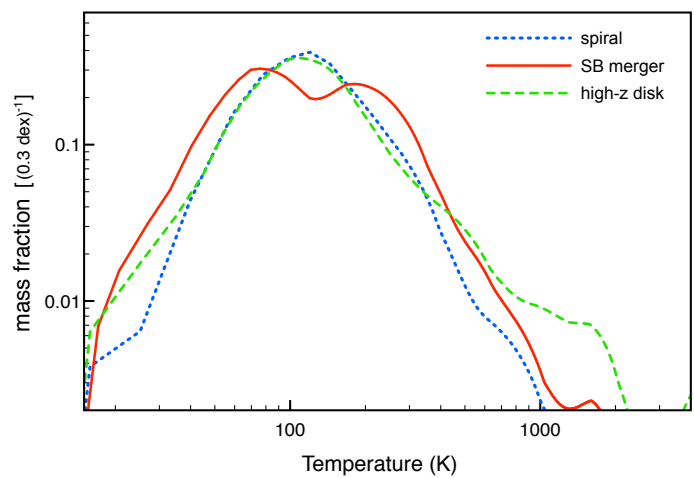}
\caption{\label{fig_Tpdf3}  Temperature PDF for the three types of galaxies, averaged over the selected snapshots. Gas denser than 100\,cm$^{-3}$ is considered here.}
\end{figure}


The average density PDF for our SB merger and spiral simulation snapshots are shown in Figure~\ref{fig-pdf}. The SB merger shows a strong excess of high-density gas with densities above $10^3$\,cm$^{-3}$ ; the mass fraction of gas found at densities in the $10^{4}$--$10^{5}$\,cm$^{-3}$ range is more than ten times larger than in spirals. This high-density gas excess is found in major merger simulations on various interaction orbits, once the numerical resolution is high enough \citep{Teyssier10, bournaud12, powell}. It arises from compressive regions in the tidal field of interacting galaxies, frequently found in early-stage interactions, inducing an excess of compressive turbulent motions down to small scales (10-100\,pc) and persisting until the final galactic coalescence. Such excesses of high-density gas are also suggested from observations of high-density tracers in starbursting mergers \citep{guo,garcia-c}.

Compared to this strong excess of high-density gas, there is no major excess of warm molecular gas in our SB mergers. The temperature PDF for molecular gas is shown in Figure~\ref{fig_Tpdf3}. Of course, increased stellar feedback from the triggered star formation activity does produce warmer gas in the SB merger model compared to spirals. However this warm gas excess\footnote{interestingly, the extra warmed gas in the SB mergers is not just from stellar feedback, but also from enhanced small-scale shock compression (Bournaud et al. in preparation)} involves less than 7\% of the total molecular gas mass : the mass of gas warmer than 100\,K in the SB merger temperature PDF in excess of the spiral temperature PDF is 5.8\% of the molecular gas mass. Actually, gas affected by stellar feedback in our SB merger simulation rapidly evolves into a low-density wind ($<10^{2-3}$\,cm$^{-3}$) representing a small fraction of the ISM mass, and not the molecular phase. Hence, the warm gas excess is much weaker than the excess of high-density gas independent of temperature. 

In order to determine the contribution of high-density and diffuse gas on the overall CO SLED, we perform control experiments with matched density or temperature distributions between spirals and SB mergers. The CO(5-4)/CO(1-0) line intensity ratio on the average CO SLED is 8.1 times higher in SB mergers than in spirals. When we re-analyze the simulations assuming that all gas denser than $10^2$\,cm$^{-2}$ is isothermal at 300\,K (for both the mergers and the spirals), this factor becomes 6.4 -- hence the difference is still largely present without any temperature effect. Conversely, when we selected gas cells in the SB merger model with a probability computed to match the density PDF of the spiral snapshots (i.e., we keep most cells with densities of a few $10^2$\,cm$^{-3}$, but only a small fraction of the cells with the highest-densities, to artificially ``remove'' the dense gas excess), without modifying the temperatures, the same ratio drops to 2.8, showing that high CO excitation in SB mergers woulds be much weaker with only a warm gas excess and no high-density gas excess.

Hence, the simulations clearly show that the differences in the CO SLED between disks and SB mergers are mainly due to the excess of dense gas created by gas compression in the mergers. This gas compression results both from the large-scale gas inflows and the compressive properties of merger-induced turbulence on small scales (e.g., Renaud et al. 2014).

\subsubsection{High-redshift disks}

\begin{figure}
\centering
\includegraphics[width=0.5\textwidth]{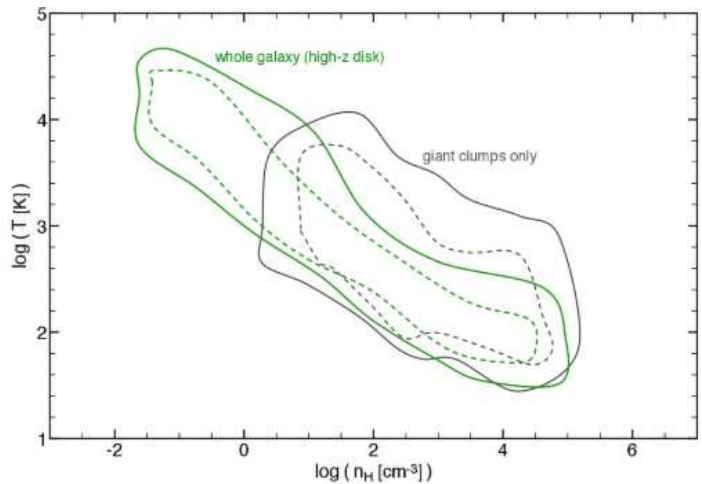}
\caption{\label{fig-n_T_deco} Phase-space distribution of the ISM in a high-z disk snapshot (black) and its giant clumps (green), selected as in Figures~\ref{fig-snapshot} and \ref{fig-sled-deco}. The figure shows the density-temperature space, with isodensity contours encompassing 90\% of the total gas mass (solid) and 67\% of the total gas mass (dashed). }
\end{figure}

\begin{figure}
\centering
\includegraphics[width=0.5\textwidth]{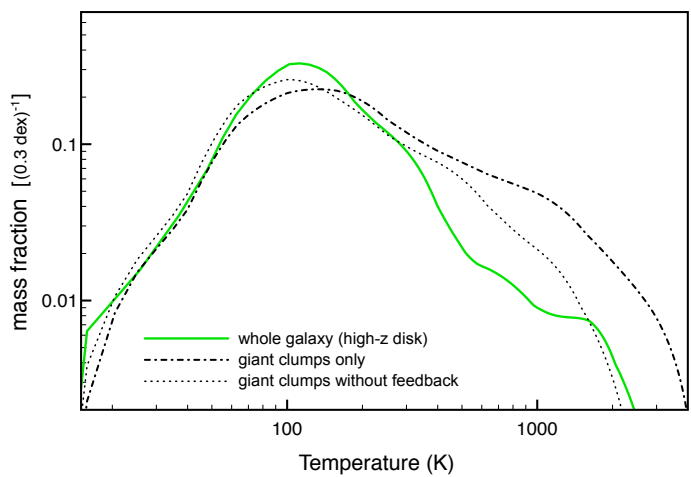}
\caption{\label{fig_Tpdf}  Temperature PDF for the high-z disk snapshot from Figures~\ref{fig-snapshot} and \ref{fig-sled-deco}, showing the temperature distribution of gas in the $10^3 - 10^4$\,cm$^{-3}$ density range (i.e. moderately dense molecular gas). The thick green line is for the entire galaxy, and thick dashed line for the giant clumps alone (selected as in Figure~\ref{fig-snapshot}) and the thin dashed line is for the giant clumps in the same simulation re-run without stellar feedback for 50\,Myr before the analyzed instant.}
\end{figure}

The phase space distribution of the interstellar gas in our high-z disk models (Figs.~7 and 8) shows a modest excess of high-density gas compared to spirals, but in much weaker proportions than in SB mergers. There is also a small excess of warm gas, somewhat larger than in SB mergers, but encompassing a small fraction of the molecular gas mass: the excess of gas at $n_H > 10^2$\,cm$^{-3}$ at $T>10^3$\,K, compared to spirals, represents 2.8\% of the mass of gas denser than $10^2$\,cm$^{-3}$. This is consistent with the fact that the average CO SLED of high-z disks are close to that of low-redshift spirals, especially when compared to starbursting mergers : the excess of warmed gas remains quite limited, and the large excess of high-density gas present in SB mergers and explaining their high excitation is mostly absent in high-z disks.

Things change at the scale of star-forming clumps, which we analyze using the same spatial selection as in Section~3.3 and Fig.~\ref{fig-snapshot}. These giant clumps span the high-density regime of the global density distribution function, but do not induce a major excess of high-density gas as that observed in SB mergers. Indeed, typical numbers for a giant clumps in a redshift two disk galaxy are a gaseous mass of $5 \times 10^8$\,M$_\odot$, an in-plane half-mass radius of 300\,pc, and a large vertical scale-height of 500\,pc or more (in both observations and hydrodynamic simulations, e.g. Bournaud et al. 2014 and references therein), yielding an average number density of 50-100\,cm$^{-3}$, only marginally higher than that of the entire host galaxy. 

However, the clumps do induce a significant excess of warmed gas at densities of 100 to 10$^5$\,cm$^{-3}$. Indeed, the non-linear dependence of star formation on gas density is such that star formation is more concentrated in the clumps than the gas mass itself (see Section~2), and the clumps gas is hence more exposed to stellar feedback per unit gas mass than the inter-clump gas. A consequence is that galactic outflows are mostly launched by the giant clumps  in our models \citep{Bournaud14, Perret14} (consistently with observations \citealt{Genzel2010}, and theory \citealt{DekelKrumholz13}). This warm gas excess is illustrated by the phase-space distribution of the clumps gas compared to that of the entire galaxies on Figure~9. While less than 10\% of the gas mass in the 100-1000\,cm$^{-3}$ density range is warmer than 10$^3$\,K in the entire high-z galaxy, this fraction becomes as high as 23\% in mass at the scale of the main giant clumps -- it is below 5\% in the inter-clump gas, and in low-redshift spirals.  

We further checked that this warm gas excess is the main caused for the bumped CO SLED found for high-z disks, by re-analyzing the simulation with a temperature capped at 300\,K for any gas denser than 100\,cm$^{-3}$. This experiment reduces the CO(4-3) line intensity to 63\% of its initial value for the entire galaxy (and to 39\% of its initial value for the giant clump regions alone), bringing it close to Milky Way-like excitation. This experiment also increased the typical \aco range from 3.9--4.5 to 4.0--4.7, again bringing it close to the values obtained for spiral disks. 

\smallskip

Thus, there is a significant difference between the high-excitation components found in SB mergers and high-z disks, regarding their physical origin. The highly excited CO SLED of SB mergers results mostly for the excess of high-density, rapidly star-forming gas among the molecular phase, which is the outcome of tidal compression and modified ISM turbulence (excess of compressive turbulent modes, Renaud et al. 2014). The moderately excited, bumped CO SLED and moderately lowered \aco factors for high-z disks results from gas heated by shocks and/or stellar feedback in the giant clumps of star-formation that are ubiquitous in high-redshift star-forming galaxies, without involving a major excess of very high-density gas. We analyze this further by examining the temperature PDF on the high-z disk simulation (Fig.~\ref{fig_Tpdf}) for the entire galaxy, for the clumps alone, and for the same model evolved without stellar feedback during 50\,Myr before the analysis. We find that the warm gas excess is largely driven by stellar feedback in the high-SFR-density clumps, but that a substantial excess of warm gas would still be found in the clump regions compared to the other regions, this part at least being attributable to (turbulent) shocks and compression of gas in the clumps.

It is interesting to note that gas heating by stellar feedback behaves differently in the SB mergers and high-z disks. These models have similar SFRs, so similar rates of energy injection through feedback. In SB mergers, less than 10\% of the mass of gas denser than 100\,cm$^{-3}$ is significantly heated (see 3.4.1 and Fig.~8) and a larger fraction of the re-injected energy is rapidly found in thermal and kinetic form in a low-density wind (Bournaud et al. in preparation). In the high-z disks, and in particular in their giant clumps, most of the feedback energy is injected in dense kpc-sized clumps and trapped in dense gas, resulting in a larger mass fraction of heated gas for densities larger than 100\,cm$^{-3}$ (about 30\% of the dense gas mass inside the giant clumps is significantly heated, see Fig.~9). In spite of the warmed component in the high-excitation clumps, the bulk of the interstellar gas mass is only modestly heated compared to nearby galaxies (Fig.~8): this is consistent with the dust temperature estimates for Main Sequence galaxies at $z$ $\approx$2 by \citet[][see also \citealt{magnelli13}, \citealt{saintonge14} and \citealt{genzel14}]{magdis}.

\subsection{Origin of the \aco variations}

We have shown above how the excess of high-density gas in SB mergers can explain the excitation of high-J CO lines. Yet such a dense gas excess does not necessarily explain the lowered \aco found in these systems. {\bf The CO(1-0) emission is largely thermalized\footnote{\bf a large fraction of the CO is at moderate densities about 100\,cm$^{-3}$ but the CO(1-0) emission emerging from the galaxies is dominated at 73\% by gas denser than 1000\,cm$^{-3}$.} }so that increased gas density cannot yield increased line temperatures at fixed H$_2$ mass. Gas warmed by feedback effect may play a role, too. To probe this potential effect, we use again the control experiment in which the SB merger snapshots and the spiral ones have matched temperatures, described in Section~4.1.1. The difference between the \aco factors of SB mergers and those of spirals is reduced only by 26\%, suggesting only a moderate contribution of the temperature effects in the variations of \aco from disks to SB mergers.

A major difference in the ISM of SB mergers is that it is much more turbulent than in progenitor spiral disks. In our models, we find that the mass-weighted average of the 1-D velocity dispersion $\left < \sigma \right >$ is 4.6 times higher in SB mergers than in spirals (here measured for densities in the 300--3000\,cm$^{-3}$ range to avoid any dependence on density). This is consistent with many merger simulations \citep[][and references therein]{bournaud10-review} and observations \citep{irwin94, elmegreen95, duc00, wei12}. In addition the turbulent velocity field in mergers show an excess of compressive (curl-free) motions compared to solenoidal (divergence-free) motions that dominate the ISM in spirals -- this outcome of the tidal galaxy interaction was presented in detail in Renaud et al. (2014). As a result, along a line-of-sight crossing dense gas cloud, the turbulent motions are more likely to be oriented radially from the cloud, along the line-of-sight, rather than curling around the cloud perpendicular to the line-of-sight: this radial compression is the source of the high-density gas excess in mergers. Indeed we measure that the $dV/dr$ velocity gradient along lines of sight crossing gas denser than 300\,cm$^{-3}$ is 7.6 times higher in SB mergers than in isolated spirals, a factor even larger than for the bulk velocity dispersion. The CO linewidths emitted by such regions in SB mergers are broadened by similar factors, and at the same time the line temperatures are affected in weaker proportions. For instance, at a density of 4000\,cm$^{-3}$ and at 50\,K, gas in spirals has a mean $dV/dr$ of 0.72\,km\,s$^{-1}$\,pc$^{-1}$ corresponding to a CO(1-0) line temperature of 42.1\,K in our LVG tables. The same gas in SB mergers has a mean $dV/dr$ of 5.1\,km\,s$^{-1}$\,pc$^{-1}$ corresponding to a CO(1-0) line temperature of 34.3\,K, and, simultaneously, a 7.1 times broader linewidth, resulting in a CO(1-0) intensity more than five times higher per gas mass in this representative regime. 
Hence the properties of ISM turbulence, modified in SB mergers compared to isolated spirals, cause a large broadening of emission lines especially on lines of sight emanating from dense clouds, increasing the intensity even when the line is thermalized, and lowering the \aco conversion factor.

\medskip

The behavior of high-redshift disks is different. Like mergers, they are highly turbulent systems, because of the gravitational instability and stirring by their giant clumps, but the high velocity dispersions are mostly found in the giant clumps: we measure an increase in the mass-weighted velocity dispersion of a factor 4.7 in the giant clumps selected in Figure~5 (compared to spirals), but only a factor 1.9 for the rest of the galaxy. The clumps dominate the high-excitation components, but have a limited contribution to the CO(1-0) line intensity. Furthermore in the case of high-redshift disks, there is no preferential triggering of compressive motions, as opposed to SB mergers. In this case, we measure that the mean $dV/dr$ velocity gradient is 2.1 times higher in high-z disks compared to spirals. The potential decrease of \aco is thus much more moderate than in SB mergers. At the same time these systems are globally more gas rich than nearby mergers with higher mean densities (Figure~7) which can lower the line temperature. The decrease of \aco is thus much more modest than in nearby SB mergers.

\bigskip

{\bf Grain photoelectric heating from UV star light and cosmic rays can be important heating processes in the low-excitation CO regions \citep[e.g.,][]{lequeux04}. Cosmic rays are neglected in our simulations, and UV stellar light heating is also neglected in molecular regions. Photoionization by stellar photons is included in the ionized HII regions (Renaud et al. 2013) and given that most HII regions are close to the resolution limit of our simulation the nearby molecular gas is numerically heated by thermal diffusion, providing a crude model of stellar feedback heating. An explicit modeling of the heating process would probably not yield more accurate results, as the stellar photons would generally be absorbed within the numerical resolution elements, especially in dense molecular regions. In any case, these processes depend mostly on the SFR, and since the SFE is about similar in nearby spirals and high-redshift disks, the average heating rate per unit gas mass should be about similar, in contrast with turbulence dissipation which largely increases from nearby spirals to high-redshift galaxies and starbursting mergers. In starbursting mergers the extra heating from SFR-dependent processes could become more important per unit gas mass and boost the high-excitation components, nevertheless in our model the extra heating in SB mergers seems generally modest (Fig.~8).}

\subsection{The case of high-redshift mergers}

High-redshift mergers were not included in the present study. The main reason is the physical uncertainty in the typical properties of high-redshift mergers. While low-redshift simulations and high-z disk ones have relatively well converged with resolution, feedback models, details of orbital parameters (Renaud et al. 2014, Bournaud et al. in preparation), and have an ISM structure very closed to observed ISM power spectra \citep{Bournaud10, Combes13}, the situation remains quite unclear for major mergers of high-redshift galaxies. Models with supernovae feedback found that they can trigger a strong elevation of the SFR, like low-z mergers, starting at a higher SFR level because of the rich gas reservoirs. But more recent models with higher resolution and thorough stellar feedback schemes actually question the ability of high-z mergers to trigger the SFR as easily and frequently as low-redshift events, as the presence of dense giant clumps in pre-merger galaxies may somewhat saturate the SFR in many cases (Perret et al. 2014). The situation is also unclear in observations, as the rate of mergers should be much higher at high redshift, but the fraction of starburst galaxies with a specific star formation rate much above the average appears to barely increase with redshift (\citealt{Rodighiero}, \citealt{sargent}, Schreiber et al. in preparation). Additionally, very high SFRs in high-redshift systems may alter the excitation process if the level populations are affected by radiation pumping. 

Another reason for not studying high-redshift mergers here is more technical: the number of resolution elements required in our AMR simulations increases when we move to high redshift and increase the total gas mass and/or model merging systems with an excess of high-density gas (requiring finer AMR cells). Combining these two situations in a high-z merger model would be extremely costly and would force us to use a lower resolution for our entire study, while the representativity of such a model, even in terms of global SFR, would be largely uncertain.

\section{Discussion}

\begin{figure}
\centering
\includegraphics[width=0.47\textwidth]{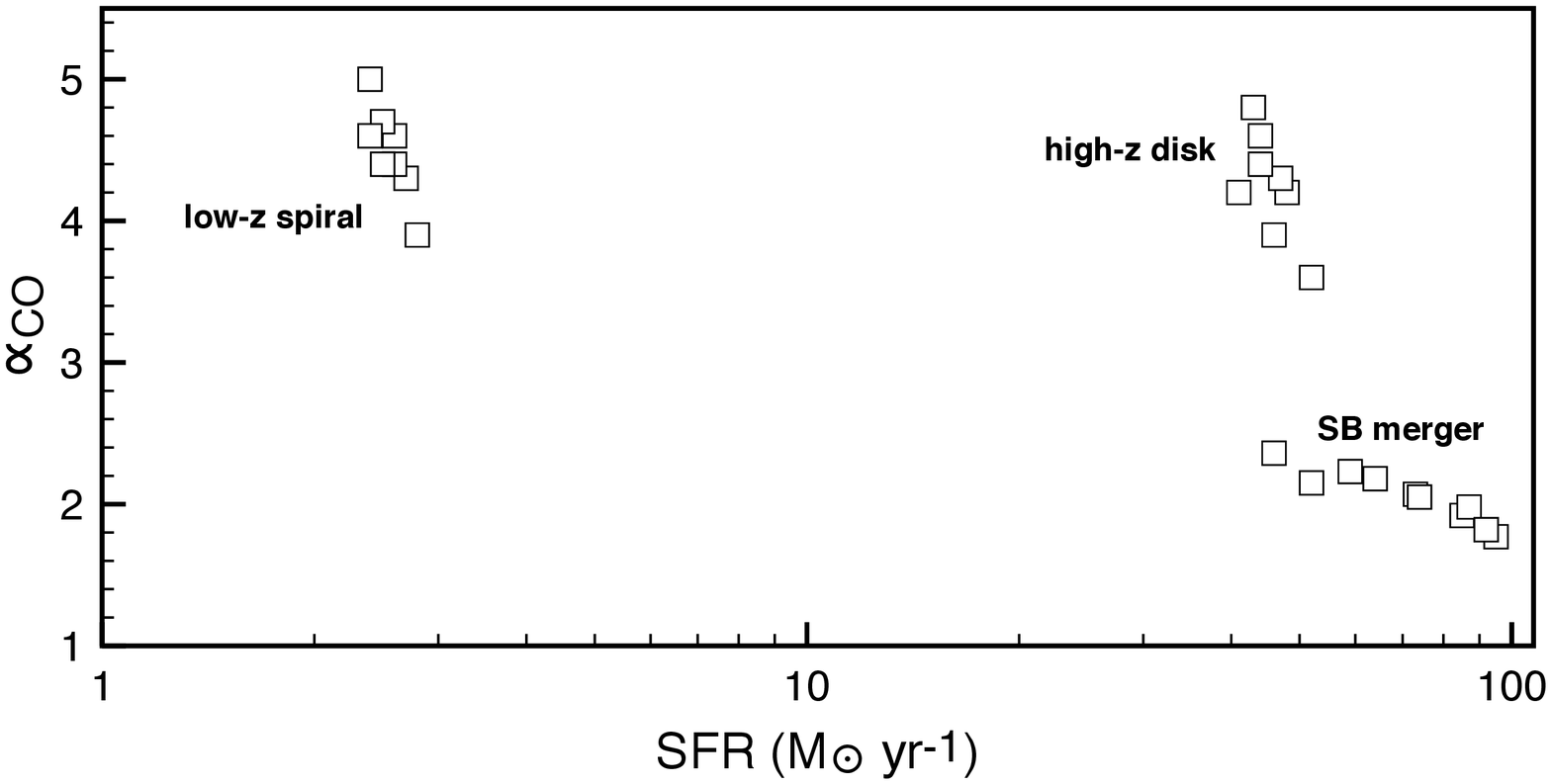}\\
\includegraphics[width=0.47\textwidth]{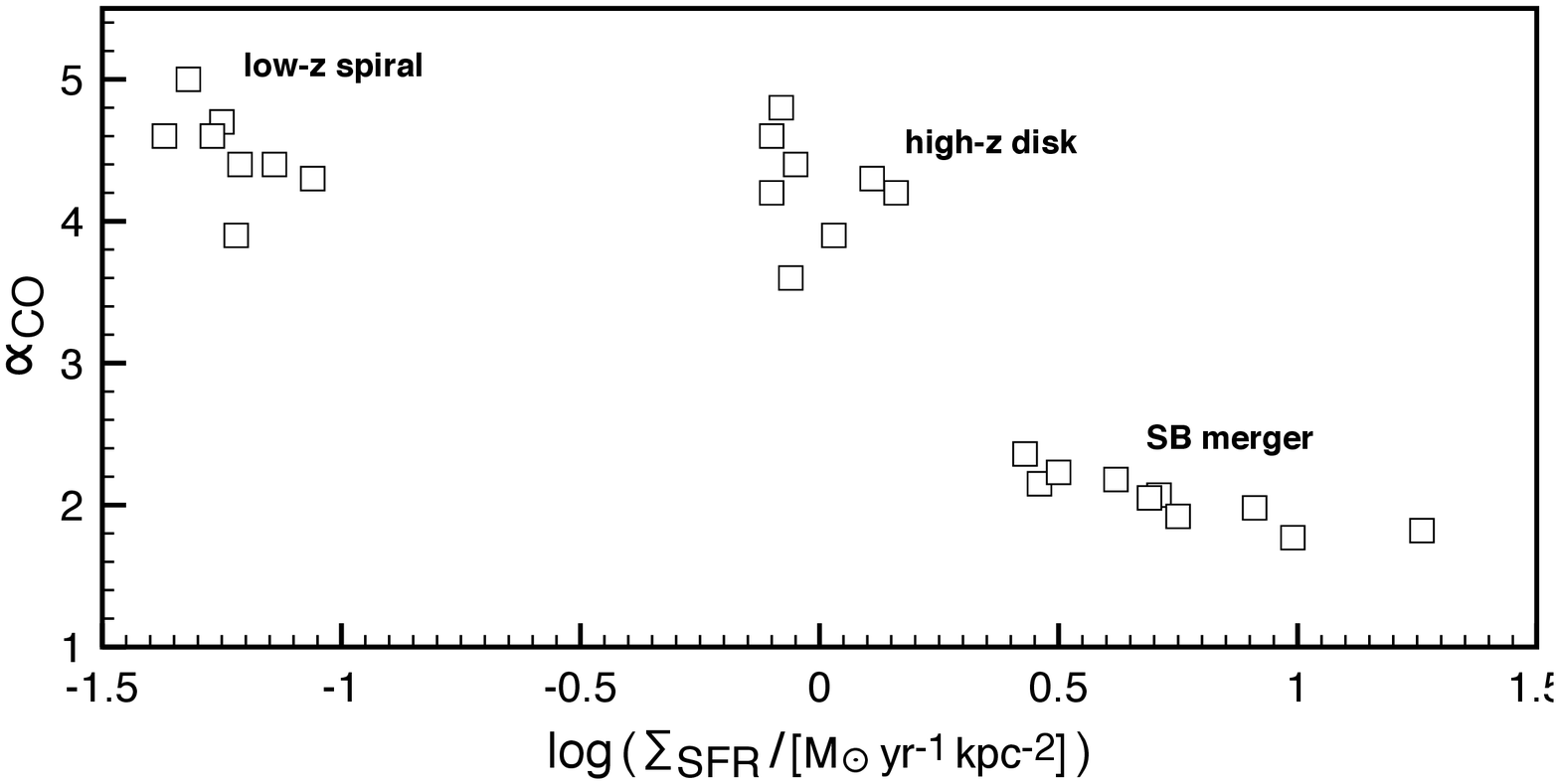}\\
\includegraphics[width=0.47\textwidth]{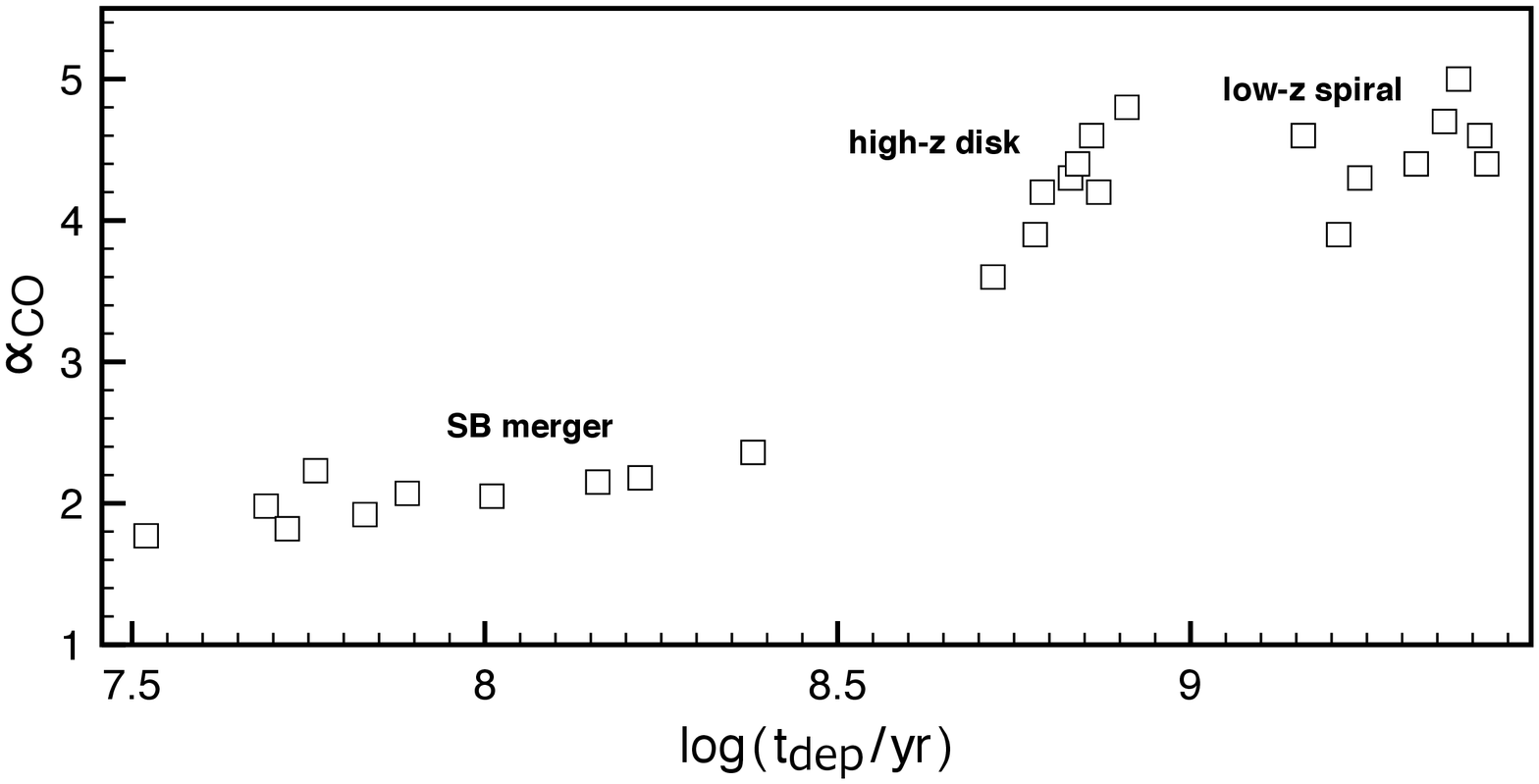}
\caption{\label{fig_compar} Comparison of the \aco conversion factor estimated for each snapshot of each galaxy type (averaged other three projections and two assumptions for the molecular H$_2$ mass in each case) to the absolute SFR (top), SFR surface density (middle) and gas depletion time (gas mass divided by SFR, bottom). }
\end{figure}

\subsection{Comparison of \aco to various parameters and models}

Our high-resolution AMR simulations coupled to LVG modeling of the CO emission and radiative transfer made it possible to predict the typical CO SLED and \aco conversion factor for various galaxy type. One of the main finding is that the CO excitation and \aco factor are not tightly linked to the SFR of a galaxy (see Fig.~\ref{fig_compar}). Although such a correlation does exist at fixed galaxy type (for instance between different snapshots of our SB merger model, see Section~3.2), the high-z disks have SFRs about as high as some SB merger snapshots, but they have an \aco factor almost as high as low-z spirals, and much higher than SB mergers (i.e. LIRGS/ULIRGs). When a high SFR (such as 50-100\,M$_\odot$\,yr$^{-1}$) results from a major merger/interaction in a brief starburst phase, the modified ISM turbulence properties result in a strong reduction of the average \aco by a large factor compared to Milky Way-like spirals -- here by a factor 2.5-3, but note that we have mostly analyzed moderately starbursting phases, so that extreme SBs may likely have stronger dense gas excesses and even lower \aco factors. In contrast, when a similarly high SFR in a galaxy results from a high gas fraction at high redshift with a long duty cycle of sustained star formation \citep{daddi-co, Elbaz11}, the \aco factor is much more moderately impacted. 

Figure~\ref{fig_compar} compares the value of \aco for all types of galaxies to the SFR, the SFR surface density $\Sigma_{\rm SFR}$ and the gas depletion time $t_{\rm dep} = M_{\rm gas}/SFR = 1/{\rm SFE}$. The two latter quantities tend to show a better correlation with \aco than SFR, but still show strong systematic deviations for at least one galaxy type compared to the two others, showing that the dependence on galaxy type found in our work cannot be rendered by such global parameters. In particular, extrapolating a trend with $t_{\rm dep}$ or SFE=1/$t_{\rm dep}$ from SN mergers to low-z spirals would lead to underestimate the typical value of \aco for high-z disks by about 30\% ($\alpha_{\rm CO}$ $\approx$3.1 instead of 4.3). Modelling the excitation of CO based on the global SFE, as done for instance in \citep{ppdpl}, could hence over-estimate the CO excitation in high-z star-forming galaxies. As we have shown, excitation of CO by stellar heating and feedback mechanism is significant only in the giant clumps of high-z galaxies, leaving the other half of the gas mass about unaffected. \\
We observe a better correlation between \aco and $\Sigma_{\rm SFR}$ in Figure~\ref{fig_compar}, although here again interpolating a power-law relation between spirals and mergers would under-estimate \aco in high-z star-forming galaxies by about 20 percent. In this sense, our results are in relatively good agreement with the model proposed recently by \citet{nara-krumholz}. The SFR density is however not the main common driver of the observed trend: the modest lowering of \aco in high-z disks does result from high gas excitation in regions of high SFR densities (giant clumps) and relatively large turbulent motions, but the large lowering of \aco in SB mergers mostly results from large line-widths in a highly turbulent ISM with ample velocity gradients dV/dr.

The retrieved ranges for \aco values in the three types of galaxies considered here are in agreement with those derived from dynamical arguments by \citet{daddi-co}. Our models thus lend further support to the use of different \aco factors for disks and for SB mergers, even when the disks have high SFRs at high redshift, and the resulting offset of starbursts in the Schmidt-Kennicutt plane (Daddi et al. 2010, Genzel et al. 2010, see also Saintonge et al. 2012). The theoretical interpretation proposed here is self-consistent, as the dense gas excess in mergers compared to gas-rich disks justifies the use of different \aco conversion factors, and at the same time this dense gas excess is theoretically expected to change the location of SB mergers in the Schmidt-Kennicutt diagram by lowering their gas consumption timescale by about 1\,dex at fixed gas surface density \citep{Renaud12}.

There are of course other possible sources of variations of the \aco factor, even at fixed galaxy type. A key one on galactic scales is probably the gas phase metallicity \citep{Israel, genzel-z}. Here we have neglected these effects, assuming solar metallicity for all our galaxies of several $10^{10}$\,M$_\odot$. Variations with redshift up to at least $z=1-2$ \citep{erb,queyrel-z,ewuyts} remain limited in this mass range: when compared to the \citet{genzel-z} relations, they may only induce variations of \aco that remain smaller than those found in the present work with galaxy type. Similarly, the plausible dilution of the gas metallicity in mergers \citep{ellison-z, dimatteo07, queyrel} is typically too small to dominate over the changes of \aco with galaxy type found in our work. Metallicity-induced variations of \aco may dominate over the effect studied here for much lower mass galaxies ($<$$10^{10}$\,M$_\odot$) or at even higher redshifts ($z$$>$2-3).

\subsection{Is excitation a good discriminator of \aco~? }

\begin{figure}
\centering
\includegraphics[width=0.47\textwidth]{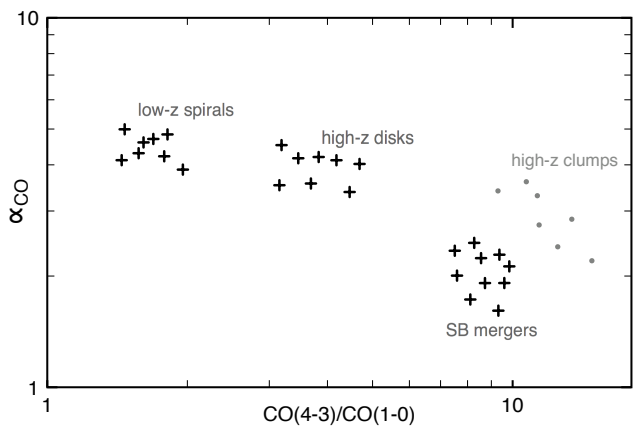}
\caption{\label{fig_excit_aco} Comparison of the CO(4-3)/CO(1-0) line intensity ratio to the \aco conversion factor, for each snapshot of each galaxy type (averaged other three projections and two assumptions for the molecular H$_2$ mass in each case), showing that excitation is not necessarily a good discrimination for \aco. Giant clumps identified in a high-z disk snapshot are also shown (dots).}
\end{figure}

\begin{figure}
\centering
\includegraphics[width=0.47\textwidth]{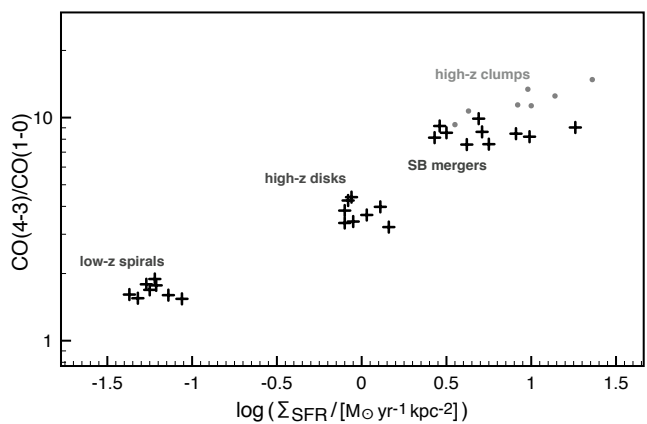}
\caption{\label{fig_excit_vs_sigma} Comparison of the CO(4-3)/CO(1-0) line intensity ratio to the surface density of the SFR, showing a much tighter correlation of CO excitation with SFR density and that of \aco with SFR density (Fig.~\ref{fig_compar}). Giant clumps identified in a high-z disk snapshot are also shown (dots).}
\end{figure}

Figure~\ref{fig_excit_aco} shows the CO(4-3)/CO(1-0) intensity ratio and \aco conversion factor for each model snapshot. Although the highest-excitation systems have the lowest \aco conversion factors, the correlation is very poor, not just because of large fluctuations but also because of systematic effects. In particular, high-redshift star-forming galaxies have high CO(4-3) luminosities coming from the highly-excited components in their giant clumps, these giant clumps represent only a limited fraction of the bulk gas mass, and only moderately lower the $\alpha_{\rm CO}$ factor of their whole host galaxies. Similar properties are found for CO(5-4)/CO(1-0) and, in weaker proportions, for CO(3-2)/CO(1-0). Our results thus indicate that CO excitation is likely not a good discriminator for the $\alpha_{\rm CO}$ conversion factor, especially when different types of galaxies are considered for the same amount of star formation (i.e., high-z disks versus starburst mergers). The situation becomes even more degenerate when high-redshift giant clumps are considered (Fig.~12) as their own CO excitation is even higher than that of SB merger, but their \aco factor remains larger -- these giant clumps have high gas densities, they are highly turbulent, but however have $dV/dr$ radial velocity gradients lower than mergers.

The behavior of CO excitation across galaxy types and redshifts indeed appears to be different from the variations of \aco. As shown on Figure~\ref{fig_excit_vs_sigma}, CO excitation is well correlated to the surface density of the SFR, with little systematic deviations for high-z disks -- in contrast to the large systematics found when comparing \aco to the surface density of the SFR (Fig.~\ref{fig_compar}, middle panel). Even the individual giant clumps inside high-z disks follow the tight trend between SFR surface density and \aco on Figure~13.

\begin{figure}
\centering
\includegraphics[width=0.24\textwidth]{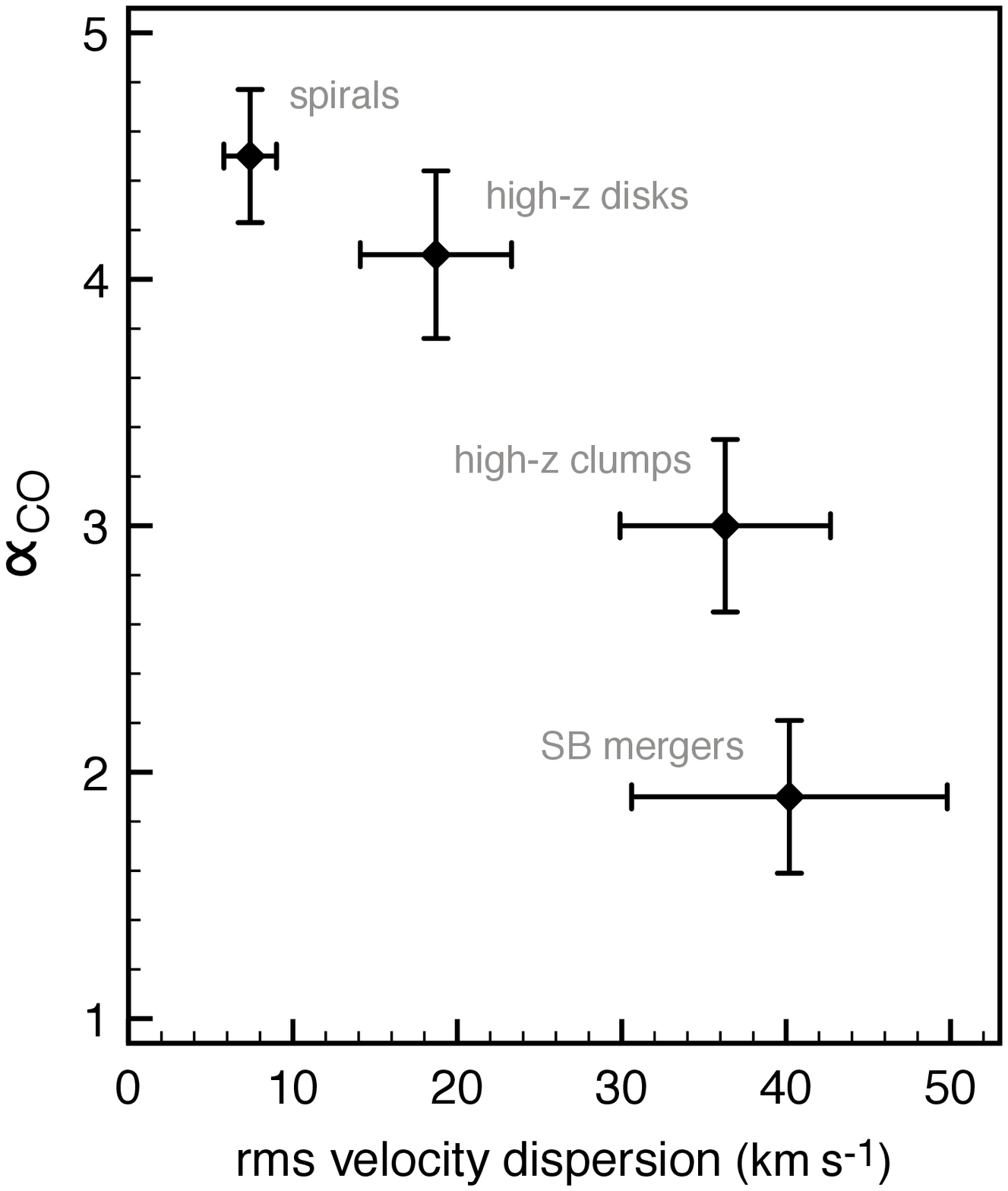}
\includegraphics[width=0.24\textwidth]{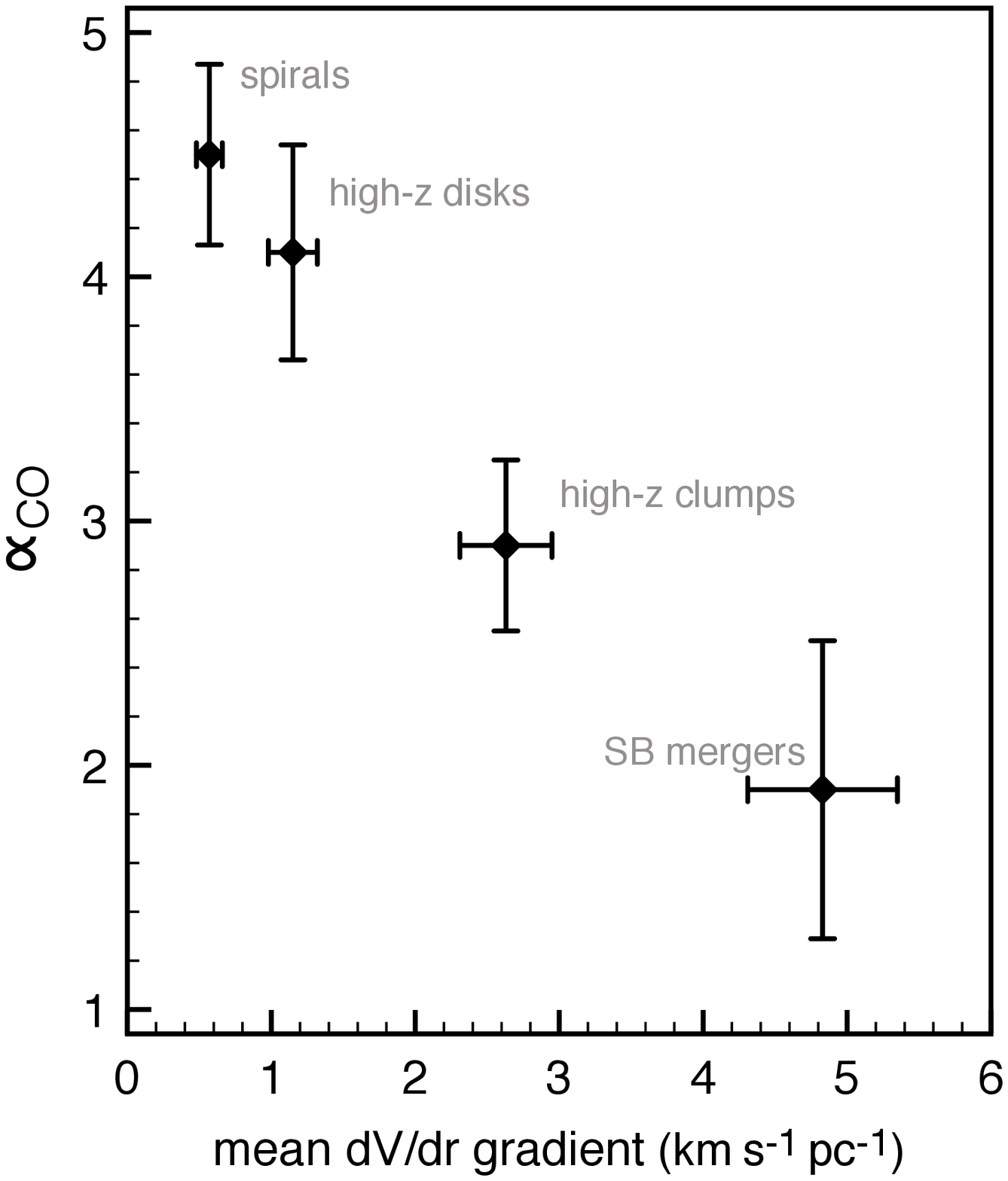}\\
\caption{\label{fig_aco_turb} Comparison of the \aco conversion factor to the bulk of ISM turbulence (left: r.m.s. 3D velocity dispersion divided by~$\sqrt(3)$, mass-weighted average of the measurements for gas identified as molecular) and to the velocity gradient along line-of-sights of molecular emission (right: mass-weighted average of the $dV/dr$ gradient along lines-of-sight crossing gas identified as molecular. Average values and r.m.s. scatter are displayed for each galaxy types, and for the giant clumps of high-z galaxies.}
\end{figure}

\subsection{Can \aco be tightly linked to ISM turbulence~?}

Our simulations and analysis have shown that the dependence of \aco on galaxy type is largely driven by the CO large line-widths in turbulent media. This could lead to the idea that measuring the turbulence of a given galaxy's ISM could be used to estimate its \aco. This would actually be largely inaccurate. As shown on Figure~\ref{fig_aco_turb} there is only a broad trend for lower \aco at high ISM turbulent velocities $\sigma$, and the clumps of high-z galaxies are almost as turbulent as SB mergers but their \aco factors remain almost 60\% higher. Hence ISM turbulence in general is not a tight tracer of \aco variations, especially when the bulk ISM turbulence $\sigma$ is measured (i.e. the r.m.s. velocity dispersion along a given line-of-sight or in three dimensions). Note that the bulk turbulent speed in our high-z disk models, about 20\,km\,s$^{-1}$, is quite consistent with what is typically observed in $z$=2 star-forming galaxies (e.g., Tacconi et al. 2010 -- the dispersions are higher in specific locations such as inside the giant clumps, and/or in the ionized gas phase, e.g. \citealt{FS11}).

A much tighter correlation is found between \aco and the mean $dV/dr$ parameter of the various physical systems (Figure~\ref{fig_aco_turb} right). Indeed this parameter controls the CO line width more directly, as it indicates how rapidly the CO lines get Doppler shifted along a given line-of-sight through radial velocity gradients around high-density peaks. A high bulk velocity dispersion $\sigma$ does not necessarily enlarge the CO lines, for instance if gas is mostly in rotational motions around the molecular density peaks (as is often found in simulations of high-z giant clumps, Ceverino et al. 2012): in such cases the radial gradients $dV/dr$ increase more slowly than $\sigma$. Starbursting mergers have the remarkable property of favoring the compressive turbulent motions, radially-oriented around density peaks, so for the same turbulent speeds $\sigma$ the $dV/dr$ gradients get larger and the lowering effect on \aco is maximized.

Qualitatively, our results can however suggests that if galaxies continue to become increasingly turbulent at higher redshift, beyond what is observed at redshift two in recent surveys, the \aco parameters may be reduced.

\subsection{Comparison to real high-redshift star-forming galaxies: total gas fraction, conversion factor, and high-excitation components}

A noticeable prediction of this modeling work is the presence of high-excitation components in the giant star-forming clumps of high-redshift galaxies, which weakly impacts the CO(1-0) and (2-1) transitions and the associated conversion factors, but is more prominent for higher-$J$ transitions, resulting in an atypical CO SLED. Although high-excitation clumps contain only a limited fraction of the total gas mass of high-z disks, our models predict that the effect could be noticeable in integrated observations of the high-$J$ lines of high-redshift gas-rich clumpy disks. We note that we have analyzed several snapshots of a single, very high-resolution simulation of a given galaxy, without exploring the variations of clumpiness intrinsically present in high-redshift galaxies \citep[perhaps induced by variations of the gas fraction, e.g.][]{salmi}: this could induce substantial variations of the intensity of high-excitation components induced by the clump, while our representative model indicates that the CO(4-3) and CO(5-4) emission lines should generally be dominated by the highly-excited clump components. Indeed, \citet{weiss05} have shown observationally in M82 that a high excitation component is coming from star-forming clumps, along with low excitation component from the diffuse disk. The analogy lends support to the idea that the importance of high-excitation components in high-z star-forming galaxies could vary from galaxy to galaxy depending on their intrinsic clumpiness.

Our model predictions are globally consistent with the first observations of high-excitation ($J_{\rm upper} \geq 4$) CO lines in Main Sequence star-forming galaxies about redshift two by \citep{daddi14}. They find that the CO(5-4) lines are brighter than expected from the CO(3-2)/CO(2-1) line ratio. In more detail, these authors find that their observed data can be fitted well by a double-component CO SLED, with the low-excitation and high-excitation component taken from our high-z disk simulation as being the clump gas and inter-clump gas (as shown on Fig.~5 and 6). They nevertheless find that in the majority of their systems (3 out of 4) the ratio of the high-excitation component to the low-excitation one has two be somewhat increased compared to our model (and unchanged for the 4th case). The best-fitting high-excitation fraction is correlated to the clumpiness. This is consistent with the idea above that we have modeled one high-z galaxies, with representative giant clumps for by violent instability in a gas-rich disk, but in detail a specific degree of clumpiness representative for a specific and somewhat arbitrary gas fraction, not covering the natural variations of this parameter. 

The gas fraction in our models (and the resulting clumpiness) may be somewhat under-estimated compared to real $z$=2 galaxies. It typically contains a 50\% total gas fraction (slowly varying with time), counting all gas phases including the atomic one. The resulting density PDF (Fig.~7) actually shows that a substantial gas fraction lies in low-density phases below 10\,cm$^{-3}$ : this phase contains on average 26\% of the total gas mass, i.e. 12\% of the baryons, while 74\% of the total gas (i.e. 38\% of the baryons) would be in the molecular one. The density distribution of the turbulent ISM indeed preserves a substantial amount of gas in moderate-density region even if the average density of the system is high enough to form molecules. This suggests that our high-z disk model has a realistic gas fraction, but on the rather gas-poor end compared to observations, and hence that the high-excitation clumps could be somewhat more prominent in average galaxies. Conversely, if real galaxies at redshift two have a molecular gas fraction of 50\% compared to the stellar mass (i.e. molecular gas mass about equal to the stellar mass -- Daddi et al. 2010, Genzel et al. 2010, see also \citealt{magdis, genzel14}) but also one fourth of their total gas mass atomic as suggested by our model, then their total gas fraction would actually be close to 60\%. In any case, the galaxy from the Daddi et al. (2014) sample with the closest clumpiness to our present high-z disk model has a quantitatively very similar CO SLED.

\section{Conclusion}

We have used hydrodynamic simulations of low-redshift spirals, starbursting (SB) mergers, and high-z disks, with stellar masses about $6\times10^{10}$\,M$_\odot$, with a very high spatial resolution (3\,pc) making it possible to finely resolve the interstellar medium structure and its interaction with a thorough set a stellar feedback sources (photo-ionization, radiation pressure, and supernovae). To model the molecular line emission from the CO molecule, we have post-processed the simulations using a set a Large Velocity Gradient grids, accounting for the local, small-scale excitation and emission of molecular transition as well as local opacity effects depending on the phase-space distribution of the gas. 

Our results highlight a strong dependence of the CO lines properties (CO SLEDs and \aco conversion factor) on galaxy type, in addition to other known dependencies (e.g. on metallicity). We found that:

\begin{enumerate}
\item Our model correctly reproduces the CO SLED properties of nearby spirals and SB mergers from the CO(1-0) transition to the CO(8-7) one at least, and the high \aco factors estimated for the Milky Way. The absence of an explicit H$_2$ formation model in the simulations only introduces uncertainties that are smaller than the intrinsic fluctuations of the CO line luminosities and of \aco for any given galaxy type.

\item SB mergers are characterized by strong turbulent compression of their ISM. This creates an excess of high-density gas among the cold molecular phase, which drives high excitation of CO -- the excess of heating from the feedback and turbulence associated to the starburst plays a smaller role. 

\item The CO SLEDs of high-z star-forming galaxies are predicted be "bumped" around the $J_{upper}=4$ transition: namely, the CO(4-3)/CO(3-2) and CO(5-4)/CO(3-2) line intensity ratios can be as large or larger than the CO(3-2)/CO(2-1) ratio, which is found for no other type of galaxies. 

\item The main source for the high-excitation components in high-z disks is an excess of warmed dense gas, found mostly in the giant star-forming clumps that are prone to shock and feedback heating, rather than an excess of high-density gas independent of temperature like in mergers. Given its origin in giant star-forming clumps, this high-excitation component should be stronger in more clumpy galaxies.

\item Strong ISM turbulence in SB mergers generates broad CO line widths with high line intensities and low \aco factors, around three times lower than spirals. The lowest values of \aco are found for the most intensely starbursting phases. 

\item High-z disks have  \aco factors that are somewhat lower than the Milky Way-like values that characterize nearby spiral models, but much closer to these than to SB mergers. The giant star-forming clumps, that are the high-excitation sites in high-z disks, have \aco factors intermediate between SB mergers and low-z spirals. 

\item The more turbulent galaxy types or regions overall tend to have broader line widths and lower \aco, but the \aco parameter cannot be simply linked to the bulk ISM turbulent speed, as other properties of ISM turbulence play a substantial role.

\end{enumerate}

\begin{acknowledgements}
We are grateful to Desika Narayanan, Mark Sargent and Anita Zanella for stimulating discussions, to Damien Chapon for help with the implementation of some post-processing modules, and to the anonymous referee for a very careful reading and useful comments. The simulations presented in this work were performed at the {\em Tres Grand Centre de Calcul} of CEA under GENCI allocations 2013-GEN 2192 and 2014-GEN2192 and on SuperMuc at the LRZ under a PRACE allocation. We acknowledge financial support from the EC through an ERC grant StG-257720 (FB, FR). 
\end{acknowledgements}

\end{document}